\documentclass[twocolumn,showpacs,preprintnumbers,amsmath,amssymb,prb,superscriptaddress]{revtex4}

\usepackage{graphicx}
\usepackage{bm}

\begin{document}

\title{Kondo effect and spin-active scattering in ferromagnet-superconductor junctions}
\author{H. Soller}
\affiliation{Institut f\"ur Theoretische Physik,
Ruprecht-Karls-Universit\"at Heidelberg,\\
 Philosophenweg 19, D-69120 Heidelberg, Germany}
\author{L. Hofstetter}
\affiliation{Department of Physics, University of Basel, Klingelbergstrasse 82, CH-4056 Basel, Switzerland}
\author{S. Csonka}
\affiliation{Department of Physics, Budapest University of Technology and Economics, Budafoki ut 6, 1111 Budapest, Hungary}
\author{A. Levy Yeyati}
\affiliation{Departamento de F\'isica T\'eorica de la Materia Condensada C-V, Universidad Aut\'onoma de Madrid, E-28049 Madrid, Spain}
\author{C. Sch\"onenberger}
\affiliation{Department of Physics, University of Basel, Klingelbergstrasse 82, CH-4056 Basel, Switzerland}
\author{A. Komnik}
\affiliation{Institut f\"ur Theoretische Physik,
Ruprecht-Karls-Universit\"at Heidelberg,\\
 Philosophenweg 19, D-69120 Heidelberg, Germany}
\date{\today}

\begin{abstract}
We study the interplay of superconducting and ferromagnetic correlations on charge transport in different geometries with a focus on both a quantum point contact as well as a quantum dot in the even and the odd state with and without spin-active scattering at the interface. In order to obtain a complete picture of the charge transport we calculate the full counting statistics in all cases and compare the results with experimental data. We show that spin-active scattering is an essential ingredient in the description of quantum point contacts. This holds also for quantum dots in an even charge state whereas it is strongly suppressed in a typical Kondo situation. We explain this feature by the strong asymmetry of the hybridisations with the quantum dot and show how Kondo peak splitting in a magnetic field can be used for spin filtering. For the quantum dot in the even state spin-active scattering allows for an explanation of the experimentally observed mini-gap feature.
\end{abstract}


\maketitle

\section{Introduction}
The quickly evolving field of spintronics has pointed to an increasing need for a complete understanding of contacts between superconductors (SCs) and spin-polarized materials such as ferromagnets (FMs). Compared to hybrids between a normal metal and a SC one not only observes the interplay of the two densities of states (DOSs) but also effects of the different spin ordering in the contacted materials. Examples of interesting transport effects are the possibility to probe the spin polarisation by Andreev reflection \cite{PhysRevB.69.140502,Soulen02101998,Meservey1994173,PhysRevB.70.054416,2010arXiv1008.1788P}, the $\pi$-junction behavior in Josephson junctions with FM interlayers\cite{PhysRevLett.86.304,PhysRevLett.86.308,RevModPhys.77.935,PhysRevLett.95.187003}, ferromagnetically induced triplet superconductivity\cite{goennenwein,bergeretr,bergeretp,birge1,trifunovic} and the interplay of these triplet pairs with magnons.\cite{PhysRevLett.99.057003,PhysRevB.76.060504}\\
Recently the interest was triggered by the observation that spin-active scattering plays an important role in SC FM interfaces \cite{PhysRevLett.102.227005} and the experimental realisation of a ferromagnet-quantum-dot-superconductor (F-QD-S) junction\cite{PhysRevLett.104.246804} which allows to study the influence and interplay of ferromagnetic and superconducting correlations on transport\cite{0957-4484-15-7-056,PhysRevB.63.094515} in the Kondo regime.\cite{KOU01}\\
So far theoretical considerations of SC FM hybrids on the one hand have mainly concentrated on the I-V characteristics\cite{PhysRevB.81.094508,PhysRevLett.74.1657,mazin:7576,PhysRevLett.81.3247} and noise properties \cite{PhysRevLett.101.257001,PhysRevB.77.064517} of quantum point contacts. On the other hand for quantum dot (QD) geometries only the spin-dependent DOS of the FM and not the additional interface properties have been considered.\cite{soller1,morten-2008-81} We show how to access the full counting statistics (FCS) for these structures including their interface properties to allow for a direct calculation of all statistical moments of the current flow.\cite{jetp1993,nazarov-2003-35}  This allows for a calculation of noise and higher order cumulants that can be experimentally observed.\cite{reulet-2003-91,PhysRevB.75.075314}\\
In this paper we investigate the FCS of SC FM quantum point contacts (QPCs) in the presence of spin-active scattering and F-QD-S junctions in odd and even charge states. In the latter system we have to keep track of the exchange field related peak splitting of the Kondo resonance. \cite{PhysRevB.72.121302,PhysRevLett.91.127203,PhysRevLett.107.176808,2011arXiv1104.3403Y} For a QD in an even charge state we show how spin-flip Andreev reflection in combination with the exchange field leads to a new characteristic subgap phenomenon.\\
The paper is organized as follows: Section \ref{sfqpc} deals with a SC FM quantum point contact without a specific consideration of the interface properties. We find in Section \ref{sfqpcsa} that a consistent interpretation of conductance spectra is only possible by considering spin-active scattering at the interface. In Section \ref{kondo} we derive an effective model for a F-QD-S device in the Kondo limit and explain why spin-active scattering does not need to be considered. The effective model allows us to demonstrate how to use the device for spin filtering in Section \ref{spin}. To obtain a complete picture of the transport properties of F-QD-S devices we also consider the even state of the QD in Section \ref{even} and show how spin-active scattering leads to a new subgap structure. We also explain the evolution of the subgap structure in a magnetic field.
\section{Ferromagnet-Superconductor Quantum Point Contact} \label{sfqpc}
As a first test system we study a QPC between a SC and a FM (SFQPC). So far the FCS of this specific arrangement have not been considered explicitely but they resemble very much the one for the normal-superconductor QPCs\cite{PhysRevLett.87.067006,soller2,PhysRevB.50.3982} since the only difference is the spin-dependent DOS. The Hamiltonian of the system reads
\begin{eqnarray}
H = H_f + H_T + H_s \label{system},
\end{eqnarray}
where $H_f$ describes the FM lead using electron field operators $\Psi_{kf\sigma}$ in the Stoner model with an exchange energy $h_{ex}$ as in [\onlinecite{melin-2004-39}]
\begin{eqnarray}
H_f &=& \sum_{k, \sigma} \epsilon_k \Psi_{kf\sigma}^+ \Psi_{kf\sigma} \nonumber\\
&& - h_{ex} \sum_k (\Psi_{kf\uparrow}^+ \Psi_{kf\uparrow} - \Psi_{kf\downarrow}^+ \Psi_{kf\downarrow}). \label{hferro}
\end{eqnarray}
For simplicity we set $\hbar = e = k_B = 1$. The FM has a fermionic flat band density of states (DOS) with asymmetry for the spin-$\uparrow$ and spin-$\downarrow$ tunneling electrons $\rho_{f\sigma} = \rho_f (1+ \sigma P)$, where $P$ is the polarisation. The superconducting lead is described by the typical BCS Hamiltonian \cite{PhysRev.108.1175} in the language of electron field operators $\Psi_{ks\sigma}$ with its characteristic energy gap $\Delta_0$
\begin{eqnarray}
H_s &=& \sum_{k,\sigma} \epsilon_k \Psi_{ks\sigma}^+ \Psi_{ks\sigma} \nonumber\\
&& + \Delta_0 \sum_k (\Psi_{ks\uparrow}^+ \Psi_{-ks\downarrow}^+ + \Psi_{-ks\downarrow} \Psi_{ks\uparrow}), \label{hsuper}
\end{eqnarray}
leading to the energy dependent DOS $\rho_s = \rho_{0s} |\omega|/\sqrt{\omega^2 - \Delta_0^2}$.\\
As in previous treatments of similar problems\cite{2010arXiv1011.3408C} we define the voltage with respect to the chemical potential of the superconducting lead $\mu_s = 0$ so that $V= - \mu_f$, where $\mu_f$ is the chemical potential of the FM. The Fermi distribution of the SC is abbreviated by $n_s$ whereas $n_{f+}$ and $n_{f-} = 1- n_{f+}(-\omega)$ refer to the electron- and hole-like Fermi distributions in the FM, respectively.\\
The local tunneling Hamiltonian, which is responsible for transfer of electrons between the leads is given by\cite{PhysRevLett.8.316}
\begin{eqnarray}
H_T = \sum_\sigma \gamma[\Psi_{s\sigma}^+(x=0) \Psi_{f\sigma}(x=0) + h.c.],
\end{eqnarray}
where $\gamma$ is the amplitude of the tunneling coupling.\\
To study the FCS we calculate the cumulant generating function (CGF) $\ln \chi(\lambda) = \sum_{n=1}^\infty \frac{(i\lambda)^n}{n!} \langle q^n\rangle$ defined as the functional generating the irreducible moments $\langle q^n \rangle$ of the charge ($q$) distribution by differentiation with respect to the counting field $\lambda$. According to [\onlinecite{PhysRevB.70.115305,PhysRevB.73.195301,nazarov}] the connection to the former Hamiltonian is given by
\begin{eqnarray}
\chi_{SF}(\lambda) = \langle T_{\cal C} \exp[-i \int_{\cal C} T^{\lambda(t)} dt]\rangle, \label{cgf}
\end{eqnarray}
where $T^{\lambda(t)}$ denotes $H_T$ after the substitution $\Psi_{f\sigma}(x=0) \rightarrow \Psi_{f\sigma}(x=0) e^{-i \lambda(t)/2}$. ${\cal C}$ is the Keldysh contour and $T_{\cal C}$ means time ordering on it. The counting field changes sign on the branches of the contour to account for a virtual (or passive) measurement of the charge being transferred.\cite{PhysRevB.80.041309} $\lambda(t)$ is nonzero only during the very long measuring time $\tau$. The different cumulants of the distribution can be obtained via differentiation of the CGF at $\lambda = 0$. The method is by now well established and has been applied to numerous quantum impurity problems (see e.g. [\onlinecite{PhysRevLett.103.136601,PhysRevB.82.121414,PhysRevB.82.165116,PhysRevLett.98.056603,PhysRevB.79.245303}]). Since the lead degrees of freedom appear quadratically in the total Hamiltonian we can calculate the CGF exactly using the Hamiltonian approach\cite{PhysRevB.54.7366}
\begin{widetext}
\begin{eqnarray}
&& \ln \chi_{SF}(\lambda) = \tau \int \frac{d\omega}{2\pi} \left(\sum_\sigma \ln \left\{ \prod_{\alpha = \pm} \{1+ T_{e\alpha \sigma} A_\alpha (\omega, \lambda)\} + T_{A2} (2n_s -1) \{(2n_s -1)[(e^{i \lambda} -1)^2 n_{f-} (1-n_{f+})  \right. \right. \nonumber\\
&& - 2 (e^{i \lambda} -1)(e^{-i \lambda} -1) n_{f-} n_{f+} + (e^{-i \lambda} -1)^2 n_{f+} (1- n_{f-})] + 2 n_s (e^{i \lambda} - 1)(e^{-i \lambda} -1) (n_{1+} -1 + n_{1-})\}\nonumber\\
&& + T_{BC} \left\{(2n_s -1)^2 (e^{i \lambda} - e^{-i \lambda})^2 [n_{f-} e^{i \lambda} + n_{f+}e^{-i \lambda} + \beta_1 (1 - \sigma P) n_s (1-n_s) (e^{i \lambda} - e^{-i \lambda})^2] \right. \nonumber\\
&& + n_s (2n_s -1) \left\{4 (n_s -1) (n_{f+} - 1 + n_{f-})(e^{i \lambda} -1- e^{-i \lambda})^2 + \sigma P\{8[(e^{i \lambda} -1)^2 n_{f+} - (e^{-i \lambda} -1)^2 n_{f-}] \right. \nonumber\\
&& - (e^{-i \lambda} -1)^3 [e^{3i \lambda} (2n_s -1) (1+ n_{f-} - n_{f+}) - (2n_s -1)(n_{f-} - n_{f+} -1) + e^{2i\lambda} (2n_s (3+ n_{f+} - n_{f-}) - 3 + 7n_{f+})\nonumber\\
&& \left. \left. \left.  - e^{i \lambda} (3+ n_{f+} + 2n_s (n_{f+} - 3 - n_{f-}) + 7n_{f-})]\right\} +2 \sigma P [\sum_{\alpha = \pm} \alpha A_\alpha (\omega, \lambda)] \right\}\theta\left(\frac{|\omega| - \Delta_0}{\Delta_0}\right)\right\}\nonumber\\
&& \left. + \ln \{1+ T_A [n_{f+}(1-n_{f-}) (e^{2i \lambda} -1) + n_{f-}(1-n_{f+}) (e^{-2i \lambda} -1)]\} \theta\left(\frac{\Delta_0 - |\omega|}{\Delta_0}\right)\right) \label{sfcgf},
\end{eqnarray}
involving the abbreviation $A_{\alpha} (\omega, \lambda) = \left[n_{f\alpha} (1-n_s) (e^{i \alpha \lambda} -1) + n_s(1-n_{f\alpha}) (e^{-i \alpha \lambda}-1)\right]$ and the effective transmission coefficients
\begin{eqnarray*}
T_{e\sigma}(\omega) &=& \frac{4\beta_1 (1+ \sigma P)}{(1+ \beta_1 (1+ \sigma P))^2- \beta_2^2(1-P)(1+P)},\\
T_{A2}(\omega) &=& \frac{4\beta_A^2 (1+ P)(1-P)}{[(1+ \beta_1 (1+ P))^2- \beta_2^2(1-P)(1+P)][(1+ \beta_1 (1 - P))^2- \beta_2^2(1-P)(1+P)]} = \frac{T_{BC}}{\beta_1} \; \mbox{and}
\end{eqnarray*}
\begin{eqnarray}
T_A(\omega) &=& \frac{4\beta_2^2 (1+ P)(1-P)}{\beta_2^4(1-P^2)^2 + \beta_2^2 (1 - P^2) [2 - \beta_1^2 (1+P)^2 -\beta_1^2 (1-P)^2] + (1 + \beta_1^2(1+P)^2)(1 + \beta_1^2 (1-P)^2)}.
\end{eqnarray}
\end{widetext}
The SC DOS enters via the transparencies $\beta_1 = \beta_n |\omega|/\sqrt{|\omega^2 - \Delta_0^2|}$ and $\beta_2 = \beta_n \Delta_0 /\sqrt{|\Delta_0^2 - \omega^2|}$ with $\beta_n = \pi^2 \rho_f \rho_{0s} \gamma^2/2$ being the tunneling rate between the contacts. The expression is valid at arbitrary temperatures taking the temperature-dependence of $\Delta_0$ into account. The difference compared to normal-superconductor QPCs (NSQPCs) is the appearance of the polarisation $P$. Consequently the result for the CGF for NSQPCs obtained in [\onlinecite{soller2}] can easily be recovered by choosing $P=0$.\\
The transmission coefficients $T_{e\sigma}$ refer to single electron transfer, while $T_{A2}$ and $T_{BC}$ describe the additional contributions from Andreev reflection above the gap and branch crossing respectively. $T_{A}$ is the transmission coefficient for Andreev reflection processes below the gap.\\
The CGF demonstrates that also in the case of SFQPCs the elementary processes of charge transfer can be identified as normal electron transfer between the electrodes above the gap and Andreev reflection processes\cite{andreev} below the gap, see Fig. \ref{sar} (c) and (a). Andreev reflection here refers to the charge transfer via an electron that is transmitted from the FM to the SC and is retroreflected as a hole.\\
However, if we compare the results we obtain for the differential conductance to the experimental results for Al/Co contacts with good transparencies\cite{PhysRevB.69.140502} we need to introduce a sizeable broadening of the BCS DOS described by a Dynes parameter $\Gamma_D$ that has to be of the order $\Gamma_D = 0.21 \Delta_0$ to obtain quantitative agreement.\cite{PhysRevLett.41.1509,PhysRevB.70.174509} Such a distortion is unexpected since the Al/Cu contacts fabricated by the same experimental procedure do not show any distorted BCS DOS.\\
This is not a problem of the Hamiltonian approach but has also been encountered when fitting I-V spectra to an extension\cite{Rodero200167} of the Blonder-Tinkham-Klapwijk (BTK) model.\cite{PhysRevB.25.4515,mazin:7576} Instead of a Dynes parameter one could introduce an effective temperature \cite{PhysRevB.72.054510} but a reliable explanation of the spectra under debate\cite{PhysRevB.70.054416,PhysRevLett.89.166603,PhysRevB.77.233304} may only be obtained by changing the model of the interface region.\cite{PhysRevB.81.094508}
\section{Ferromagnet-superconductor quantum point contact with spin-active scattering} \label{sfqpcsa}
\begin{figure}
\includegraphics[width=7cm]{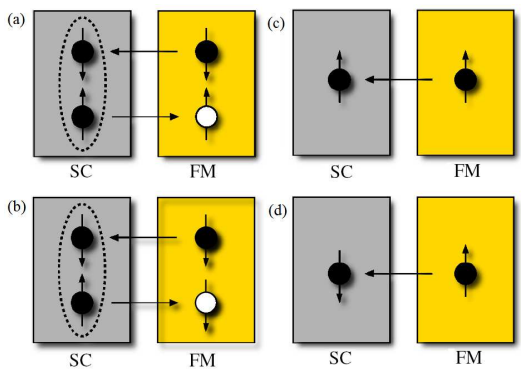}
\caption{Different transport processes in a SFQPC with spin-active scattering: in (a) we show the typical Andreev reflection and in (c) we show the typical single electron transmission between a SC and a FM. In (b) we show the SAR process involving a spin-flip at the interface and giving rise to triplet correlations in the FM. Likewise also spin-flip transmissions occur as indicated in (d).}
\label{sar}
\end{figure}

\begin{figure}
\includegraphics[width=6cm]{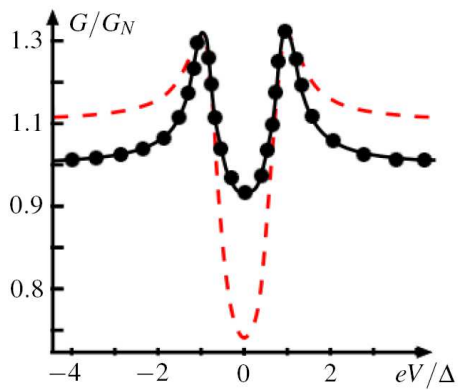}
\caption{Experimental data for the differential conductance as a function of $V$ through an SFQPC taken from [\onlinecite{PhysRevB.69.140502}] (black solid curve). The data has been normalised with respect to the normal state conductance $G_N$. Additionally, the experimental data has been scaled with the fit parameter $\Delta$ from the BTK model that is identical to the fit parameter $\Delta_0$ from our model with spin-active scattering. We plot the prediction by this model (black dots curve) for $T=0.1\Delta_0$, $\beta_n = 0.095$, $\beta_f = 0.065$, $P=0.38$ and a broadening of the BCS DOS described by $\Gamma_D = 0.09\Delta_0$. One observes the characteristic double peak structure at the superconducting gap. We also compare this result to the best possible fit without spin-active scattering using $T=0.13\tilde{\Delta}_0$, $\beta_n = 0.13$, $\beta_f = 0$, $P=0.38$, $\Gamma_D = 0.09 \tilde{\Delta}_0$ and $\tilde{\Delta}_0 = 0.75 \Delta$ (dashed red curve). $\tilde{\Delta}_0$ is the fit parameter for the gap in the model without spin-active scattering.}
\label{fig2}
\end{figure}

In the previous Section it is mentioned that a realistic description of the interface region is necessary for a complete understanding of SFQPCs. This can be achieved by considering a more complex model that explicitely includes a spin-dependent scattering potential\cite{PhysRevB.38.8823,PhysRevB.81.094508} at the interface. This is of special importance when dealing with the experimentally relevant case of strong spin polarisation $P\approx0.2-0.8$. The mechanism of spin-active scattering at the interface is the interplay of the ferromagnetic exchange field in both the bulk and the interface. In the simplest case the two magnetic moments deep in the bulk and at the interface would just be parallel. However, manifold processes may lead to an interface magnetic moment different from the bulk like using a thin magnetic layer, spin-orbit coupling, magnetic anisotropy or spin-relaxation. So far the study of the I-V characteristics of point contact spectra have been performed using a quasi-classical Green's function approach\cite{PhysRevB.81.094508,PhysRevB.80.184511} or a wave-function matching technique.\cite{PhysRevB.83.054513} Also noise properties have been analysed using a scattering states description.\cite{PhysRevLett.101.257001,PhysRevB.77.064517}\\
However, an investigation of the FCS of such setups is missing. Still, it is needed for an unambiguous identification of the charge transfer processes. In order to proceed as in Section \ref{sfqpc} we want to take a different approach compared to the quasiclassical scattering theory by following [\onlinecite{PhysRevLett.89.286803,PhysRevLett.90.116602,yamada}], where spin-active scattering is described by the introduction of an additional spin-flip contribution to the Hamiltonian
\begin{eqnarray}
H_{T2} = \sum_\sigma \gamma_2 [\Psi_{s\sigma}^+(x=0) \Psi_{f-\sigma}(x=0) + h.c.] \label{spinflip}.
\end{eqnarray}
Adding $H_{T2}$ to the system's Hamiltonian in Eq. (\ref{system}) we need to introduce a second contribution to Eq. (\ref{cgf}) to access the CGF
\begin{eqnarray}
\chi_{SFa}(\lambda) = \langle T_{\cal C} \exp [-i \int_{\cal C} dt (T^{\lambda(t)} + T_2^{\lambda(t)})]\rangle, \label{cgf2}
\end{eqnarray}
where $T_2^{\lambda(t)}$ denotes $H_{T2}$ with the additional substitution $\Psi_{f\sigma}(x=0) \rightarrow \Psi_{f\sigma}(x=0) e^{-i \lambda(t)/2}$. This allows us to calculate the CGF. The actual form of the full CGF is quite complicated due to the presence of Andreev reflection and branch crossing above the gap. We therefore only give a simplified form in Appendix \ref{appA}, Eq. (\ref{spincgf}), that allows for a clearer identification of the relevant charge transport processes above and below the gap.\\
Even in the simplified form we observe a more complicated structure of the CGF compared to Eq. (\ref{sfcgf}) since we do not only need to introduce a spin-dependent DOS but also two contact transparencies $\beta_n = \rho_f \rho_{0s} \gamma^2 \pi^2 /2, \; \beta_{f} = \rho_f \rho_{0s} \gamma_2^2 \pi^2/2$ that refer to the normal and spin-flip transparency, respectively. Consequently the CGF shows single-electron transmission for the different spins as well as spin-flip transmission processes for energies above the gap, see Fig. \ref{sar} (c) and (d). For energies below the gap we identify two types of Andreev reflection: spin-symmetric Andreev reflection (AR) and anomalous\cite{beri} or spin-flip Andreev reflection (SAR), the latter involving a spin-flip process during the Andreev reflection, see Fig. \ref{sar} (a) and (b). We therefore have obtained the FCS of all charge transfer processes that have also been identified in the quasi-classical Green's function calculation.\cite{PhysRevB.81.094508} The only difference is the description of spin-active scattering. Grein et al. use the spin-mixing angle $\theta_s$ as the phenomenological parameter whereas we use a second tunneling transparency to account for spin flips. Both descriptions are related since both give rise to Andreev bound states characterised by $T_A(\epsilon_{\pm})=1$ from which one can calculate $\theta_s$ via $\epsilon_{\pm} = \pm \Delta \cos(\theta_s/2)$.\cite{PhysRevB.81.094508}
\\
The SAR implies a spin-flip to convert a singlet Cooper pair from the SC to a triplet pair in the FM, whereas AR just transfers singlets to the FM. Spin-active scattering therefore gives rise to triplet correlations in the FM\cite{PhysRevLett.90.137003,PhysRevLett.98.077003}, which are responsible for a long-range proximity effect also in FMs.\\
Apart from the quasi-classical Green's function formalism\cite{PhysRevB.81.094508} and the approach presented here a third theoretical treatment has been frequently used for the analysis of SFQPCs: the extended BTK model.\cite{PhysRevB.69.140502} In this model the ratio of $\gamma_2$ and $\gamma$ is fixed for every possible value of $\gamma$. This model has been very successful for certain setups\cite{PhysRevB.69.140502} but fails for others.\cite{PhysRevB.80.184511,PhysRevB.77.064517,PhysRevB.81.094508} The problems in fitting conductance spectra with the extended BTK model have also been adressed before by explicit comparison to experimental data.\cite{2010arXiv1008.1788P}\\
Here we show that our model (as the quasi-classical Green's function formalism) reproduces the experimental data from [\onlinecite{PhysRevB.69.140502}] for a finite spin-flip amplitude. As the data can also be fitted by the extended BTK model this shows that one can reproduce its results.\\
We calculate the differential conductance from the current given by the first derivative of the CGF with respect to the counting field $I_{SFa} = -i/\tau \partial \ln \chi_{SFa}(\lambda)/\partial \lambda|_{\lambda = 0}$. We compare the differential conductance $dI_{SFa}/dV$ to the experimental data for Al/Co contacts.\cite{PhysRevB.69.140502} The result is shown in Fig. \ref{fig2}.
\\
We obtain perfect agreement for a reasonable Dynes parameter (meaning much smaller than the gap), which again signifies the importance of spin-active scattering for strongly polarized FMs as cobalt. The result incorporating spin-active scattering may also be compared to the best possible fit without spin-active scattering (see Fig. \ref{fig2}). We see that a consistent explanation of the experimental data heavily relies on the inclusion of a complete description of the interface region. The possibility of SAR opens a new transport channel below the gap that leads to an increased conductance for $V < \Delta_0$. Indeed the contribution by SAR is sizeable since the ratio of SAR vs. AR is determined by the ratio of $\beta_f$ and $\beta_n$ being of the order of $0.7$. This is in accordance with other studies of different point contacts.\cite{PhysRevB.81.094508,2010arXiv1012.3867H}\\
To account for the scaling of the experimental data we express our fit parameter $\Delta_0$ in terms of the BTK fit parameter $\Delta$. Both fitting parameters are only identical when we use the CGF including spin-active scattering. Also, the experiments on Al/Co contacts use samples with several transmission channels. The theoretical investigation in [\onlinecite{PhysRevB.69.140502}], however, revealed that the samples may be effectively described by a single-channel model using an effective transmission coefficient.

The FCS allows for the calculation of the noise power given by the second derivative of the CGF $S_{SFa} = - 1/\tau \partial^2 \ln \chi_{SFa}(\lambda)/\partial \lambda^2|_{\lambda = 0}$. 
In the case of small interface transparency we find the smooth transition from Andreev reflection noise (corresponding to a Fano factor of 2) below the gap to single electron noise above the gap (corresponding to a Fano factor of 1), since these represent the only dominant charge transfer processes in these regimes. This picture is very similar to the one obtained for NSQPCs since the charge transfer processes in both systems are the same. 

\section{Ferromagnet-Quantum Dot-Superconductor Device in the Kondo limit} \label{kondo}
\begin{figure}
\includegraphics[width=7cm]{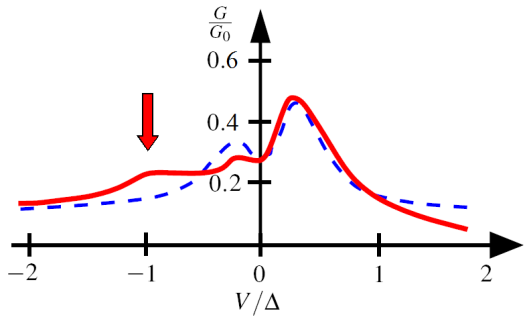}
\caption{Theoretical differential differential conductance (dashed curve) through the F-QD-S junction for $T=0.13\Delta$, $\beta_n = 0.036$, $\Gamma_K = 0.29\Delta$, $P=0.46$, $\delta_{split} = 0.16\Delta$, $T_{g} = 0.035$ and $\Delta_0 = 0.23\Delta$. One observes the characteristic double peak structure\cite{0957-4484-15-7-056}, however, now with the asymmetry related to the Kondo peak splitting. We compare our prediction to the experimental data (solid curve) taken from [\onlinecite{PhysRevLett.104.246804}] at the background voltage $V_{BG} = 1.28 V$ and $\Delta=0.14$meV. The red arrow indicates the bare SC gap observable in the experimental data.}
\label{fig4}
\end{figure}

In this Section we study the generalisation of our results to QD geometries. Even in the simplest case this requires the solution of the Anderson impurity model with ferromagnetic leads. However, for the here analysed F-QD-S experiment\cite{PhysRevLett.104.246804} the situation is such that we are clearly in the Kondo regime for an odd number of electrons on the QD.\cite{schoene2} In this case we observe the many-body spin 1/2 Kondo resonance\cite{goldhaber} for temperatures below the Kondo temperature $T_K$. Depending on the ratio $T_K/\Delta$, two scenarios may occur: for small $T_K/\Delta$ the Kondo resonance is weakly coupled to the SC due to the absence of mobile electrons at the Fermi edge. For large $T_K/\Delta$ the Kondo resonance couples to the quasiparticles in the SC. This picture is confirmed both theoretically and experimentally: theoretically the dot spectral density for a normal-QD-SC system has been analysed in [\onlinecite{PhysRevB.63.094515}] showing a cross-over from a double peak around the Fermi energy to a single Kondo resonance upon increasing onsite interaction $U$. It was also found that the double peak is due to the SC proximity effect. Experimentally for two superconducting drains and $T_K \lesssim \Delta_0$ the Kondo effect is suppressed\cite{PhysRevLett.89.256801} whereas for a SC hybrid junction and $T_K \lesssim \Delta_0$ one observes a strong suppression of the effective hybridisation between the dot and the superconducting drain.\cite{0957-4484-15-7-056}\\
This picture allows for a ad hoc but practicable approach to our more complicated setup, which is supported by comparison to the experimental data. Since we are clearly in the Kondo regime we may assume to be at the strong-coupling fixed point where a perfectly transmitting channel opens up and the transport properties of the Kondo effect can be described as the ones of a pure resonant level system as far as the electronic transport is concerned.\cite{aleiner} Additionally, we want to assume to have $T_K/\Delta$ small. Our approach will therefore be limited to the constellation of parameters $T_K < \Delta < U$. In this situation we will encounter a weak coupling of the SC to the Kondo resonance and we may use an effective description of the Kondo resonance as a resonant level weakly coupled to the SC on top of a (small) background conductance\cite{springerlink:10.1007/BF00654541} in order to describe the experiment. The background conductance describes in a first approximation the DOS outside the Kondo peak, e.g. the Hubbard subbands at $\pm U/2$. An effective model for the normal-QD-SC case was analysed in [\onlinecite{soller2,0957-4484-15-7-056}] where the resulting transmission coefficients for the Kondo resonance are just the product of the transmission coefficient for the tunneling case with the DOS of a resonant level. This simple model has been verified both theoretically and experimentally: We compared the conductance features of our approach to the theoretical calculation in [\onlinecite{PhysRevB.63.094515}] and observed good agreement in the considered range of parameters. In [\onlinecite{soller2}] the results of this ansatz were compared to the experimental data from [\onlinecite{0957-4484-15-7-056}]. Such procedure may also be applied to the FM-case as one can see from the calculation for the resonant level.\cite{soller1}\\
Addtionally we have to take into account that the Kondo resonance splits into a doublet in a magnetic field according to the Zeeman energy and is also strongly affected by the exchange field of the FM. Using Haldane's scaling method for a flat band structure with spin dependent tunneling rates and including a finite Stoner splitting of the leads an analytical formula for the energy splitting of the spin-$\uparrow$ and spin-$\downarrow$ bands is found\cite{PhysRevB.72.121302,PhysRevLett.91.127203} to be\footnote{The assumption of a flat band structure is not critical at this point since also the SC DOS is linear for energies $\omega \gg \Delta$ that are relevant for the scaling argument given here. We will substantiate this argument later.}
\begin{eqnarray}
\delta_{split} = g\mu_B B + \Delta_s + \frac{P\Gamma_K}{\pi} \ln \left(\frac{|\Delta_d|}{|U+ \Delta_d|}\right), \label{split}
\end{eqnarray}
where $\Gamma_K$ is the hybridisation of the ferromagnetic drain with the Kondo dot and $\Delta_d$ is the position of the energy level of the QD. $g \mu_B B$ is the Zeeman splitting with Bohr's magneton $\mu_B$ and the gyromagnetic ratio $g$. $\Delta_s$ is a Stoner splitting induced shift.\cite{PhysRevB.76.045321} Eq. (\ref{split}) is supported by a refined analysis based on numerical renormalisation group calculations.\cite{PhysRevB.72.121302,PhysRevLett.91.127203} The two spin bands refer to two Kondo singlets which can be described by an effective DOS with a Lorentzian shape given by
\begin{eqnarray}
\rho_{K\sigma}(\omega) = \frac{\Gamma_K^2}{\left(\omega - eV + \sigma \delta_{split}\right)^2 + \Gamma_K^2}. \label{effdos}
\end{eqnarray}
This form assumes that we have two spin split Kondo singlets that both lead to perfectly transmitting channels and are associated with the separate spin species. In the case of a FM lead spin symmetry is broken so that the Kondo screening clouds associated to the two Kondo singlets are different and thus so are the couplings of the SC to the Kondo peaks. In a first approximation they are given by the tunnel couplings in the SFQPC. The transmission coefficients for the case of a F-QD-S device can thus be deduced from the transmission coefficients for the SFQPC and the relevant effective DOS
\begin{eqnarray*}
T_{eK\sigma}(\omega) &=& T_{e\sigma}(\omega) \rho_{K\sigma}(\omega), \\
T_{AK}(\omega) &=& T_A(\omega) \rho_{K\sigma}(\omega) \rho_{K-\sigma}(-\omega).
\end{eqnarray*}
Additionally we have to include the background conductance, which may be done using the standard Levitov-Lesovik formula\cite{levitov-1996-37}
\begin{eqnarray*}
\ln \chi_g(\lambda, \tau) &=& 2 \tau \int \frac{d\omega}{2\pi} \ln \{1+ T_g [(e^{i \lambda} -1) n_{f+} (1-n_s) \\
&& + (e^{-i \lambda} -1) n_s (1-n_{f+})]\},
\end{eqnarray*}
with an energy-independent transmission coefficient $T_g$ for electron transfer. The CGF for the F-QD-S junction is given by
\begin{eqnarray}
\ln \chi_{F-QD-S} = \ln \chi_{res} + \ln \chi_g, \label{cgffqs}
\end{eqnarray}
where $\chi_{res}$ can be derived from $\chi_{SF}$ by replacing $T_{e\sigma}$ by $T_{eK\sigma}$ and $T_A$ by $T_{AK}$. In principle, one would also have to take into account branch crossing and Andreev reflection above the gap. However, we checked that in the limit of $\beta_n \ll 1$, that represents the typical experimental case of small hybridisation of the SC with the Kondo resonance, the corresponding transmission coefficients $T_{A2}$ and $T_{BC}$ may safely be neglected since their contribution is marginal.\\
We compare our results to the experimental data obtained in [\onlinecite{PhysRevLett.104.246804}]. In the experimental setup a ferromagnetic drain is formed by a Ni/Co/Pd trilayer and a Ti/Al bilayer is used as a superconducting drain. A QD forms in an InAs nanowire segment contacted by the FM and the SC. In agreement with previous experiments\cite{doi:10.1021/nl802418w,PhysRevB.74.233304} the QD is perfectly controllable by a backgate voltage. The choice of InAs is essential as its $g$-factor in a wire geometry is comparable to the (rather big) bulk value.\cite{doi:10.1021/nl802418w,2011arXiv1105.1462S}\\
Within this experiment the observed splitting of the Kondo resonance according to Eq. (\ref{split}) has been verified. To test our model for the transport characteristics we choose a charge state that exhibits a clear signature of ferromagnetic correlations, i.e. the Kondo resonance has a finite and roughly constant splitting at $B= 0T$. In this case $\delta_{split}$ is constant. We calculate the differential conductance and show the comparison of theory and experiment in Fig. \ref{fig4}.
\\
We observe acceptable agreement in the voltage range considered here. Especially we see that our model correctly describes the asymmetry of the two Kondo conductance peaks that may be traced back to the different DOS for the two spin species. Concerning the fitting parameters of our model we find that we do not observe any quasiparticle-lifetime broadening of the SC DOS and the value for the polarisation is typical for cobalt based junctions.\cite{PhysRevB.69.140502} The width of the Kondo resonance has a typical size also found in other experimental setups and theoretical treatments.\cite{0957-4484-15-7-056} From the fit we find $\Gamma_K \lesssim \Delta$, and thus $T_K \lesssim \Delta$ so that we access the interesting Kondo regime where the Kondo effect and superconductivity are concurring phenomena. We should also add that the value $\delta_{split}$ allows us to calculate the $g$-factor for the considered charge state considered since the critical magnetic $B_c$ field is related to the exchange field splitting via $g \mu_B B_c = 2 \delta_{split}$. $B_c$ has been measured in the experiment to be $64$mT. The corresponding $g$-factor would be 12, which is in perfect accordance with previous experimental studies of InAs nanowires.\cite{doi:10.1021/nl802418w} The small value $\Delta_0 = 0.23\Delta$ signifies that the Kondo resonance couples to quasiparticle states within the superconducting gap, which can be ascribed to the granularity of the metallic contacts\cite{PhysRevLett.53.2437} and/or a nonzero DOS in the nanowire sections adjacent to the QD.\cite{doi:10.1021/nl801454k} The most important source of deviations from our model is that we neglected a possible energy dependence of the background DOS and the superconducting correlations on the quantum dot. Indeed due to the latter assumption in our model we do not see the bare superconducting gap at $V=-\Delta$ (see red arrow in Fig. \ref{fig4}).\\
The result reveals two basic facets of F-QD-S junctions. The first observation is that the background DOS (given by the transmission coefficient $T_g$) is very small. The second intriguing feature is the absence of spin-active scattering. In our model we did not include a spin-active tunneling term as in Eq. (\ref{spinflip}). Such a term would couple the tunnel transmission for one spin species to the Kondo singlet for the opposite spin, which would reduce the asymmetry in the peaks. Additionally SAR processes would have to be taken into account that couple only to one Kondo singlet and would therefore lead to a pronounced subgap feature. A subgap feature of this type will be discussed in Section \ref{even}, where we show its relevance for a QD in the even state. Both characteristics are not observed and the value for the polarisation ($P=0.46$) that reflects the asymmetry of the peaks is in perfect accordance with previous experimental studies of point contacts. The absence of spin-active scattering even in the presence of a strongly polarized FM is related to the strong asymmetry of the couplings between the dot and the FM or the SC respectively. The Kondo effect is mainly due to hybridized FM bulk states so that specifics of the interface or the SC are hardly seen. This also explains why the theory in [\onlinecite{PhysRevLett.91.127203}] applies also for the case of a F-QD-S junction even though it has been derived for a QD coupled to two ferromagnetic leads.\\
The CGF in Eq. (\ref{cgffqs}) also includes the possibility of Andreev reflection. Due to the low tunneling coupling of the SC to the QD it is strongly suppressed. However, in [\onlinecite{soller2}] it was found that the presence of Andreev reflection for a normal-QD-SC junction can be decided by a noise measurement. In the case of a F-QD-S junction the Fano factor does not change considerably since Andreev reflections are not only suppressed by the small tunneling coupling but also by the Kondo peak splitting. Therefore higher order cumulants are necessary to decide the presence of Andreev reflections in these devices.\\
It is remarkable that the asymmetry and splitting deduced from the model can be explained with a reasonable choice of the $g$-factor of InAs and the polarisation of cobalt. Additionally, the asymmetry of the Kondo conductance peaks can be traced back to the different spin species which we want to exploit in the next section.
\section{Spin measurement} \label{spin}
\begin{figure}[ht]
\includegraphics[width=7cm]{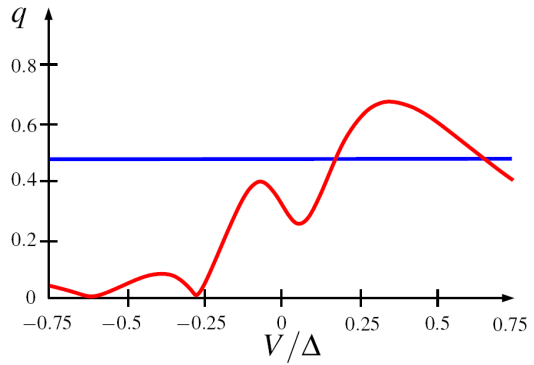}
\caption{Calculation of the quality factor in Eq. (\ref{quality}) for spin filtering with the same experimental parameters as used in the fit in Fig.~\ref{fig4}. We have neglected Andreev reflections since we have shown that they do not considerably change the conductance properties. The quality factor reaches about 70\% taking into account the effect of finite temperature in the experiment. The blue line indicates the quality factor $q= P$ of a simple tunneling junction to a FM with equal polarisation $P=0.46$.}
\label{spinfilter}
\end{figure}

In this Section we show that the F-QD-S device can be used for spin measurement. We take advantage of the above observation that Andreev reflection can be neglected as far as conductance is concerned and obtain the FCS for the separate spins as
\begin{eqnarray}
\ln \chi_{F-QD-S\sigma} = \ln \chi_{res\sigma} + \ln \chi_{b\sigma}.
\end{eqnarray}
$\ln \chi_{res\sigma}$ is obtained from $\chi_{res}$ by setting $T_{eK-\sigma} = 0$ and $T_{AK} = 0$. Likewise $\ln \chi_{b\sigma}$ is obtained from $\ln\chi_{b}$ by setting $\ln \chi_{b\sigma} = 1/2 \ln \chi_b$ since the background is assumed to be spin symmetric. Now we may calculate the conductance $G_\sigma$ for the two spin species as usual from the respective FCS. This allows us to derive the quality factor for spin filtering in our device along the lines of [\onlinecite{PhysRevB.81.075110}]
\begin{eqnarray}
q = \left|\frac{G_\uparrow - G_\downarrow}{G_{\uparrow} + G_\downarrow}\right|. \label{quality}
\end{eqnarray}
The result is given in Fig. \ref{spinfilter} using the parameters $T=0.13\Delta$, $\beta_n = 0.036$, $\Gamma_K = 0.29\Delta$, $P=0.46$, $\delta_{split} = 0.16\Delta$, $T_{g} = 0.035$ and $\Delta_0 = 0.23\Delta$ as determinded from the experimental data above. The quality factor reaches about 70\% for voltages around $0.3 \Delta$, where the conductance for the majority spin (spin-$\uparrow$) is dominant. For $V/\Delta \approx -0.15$ the minority spin component (spin-$\downarrow$) is dominant. For $V/\Delta \approx -0.3$ both spin-directions have roughly the same transmission probability ($q$ goes to zero) and for even lower voltages the spin-$\uparrow$ component again takes over, which causes another dip in the $q$ plot in Fig. \ref{spinfilter}. A possible quality factor of 70\% is much better than with a simple ferromagnetic tunneling contact as there one could only reach a quality factor equal to the polarisation $P$ meaning $\approx 46$ \%. We should emphasize that in comparison to the QD spin valve considered in [\onlinecite{0295-5075-91-4-47004}] our geometry is simpler since we do not have to work with three leads. Also we do not have to rely on interaction effects in quantum wires \cite{PhysRevB.79.085420,PhysRevLett.104.076401} or the antiresonance in double QDs. \cite{PhysRevB.81.075110}
\\
The behavior shown in Fig. \ref{spinfilter} can be explained by the interplay of the two Kondo resonance peaks that correspond to the separate spin components of the current. For $V/\Delta \approx 0.3$ electronic transport proceeds mainly through the Kondo singlet for spin-$\uparrow$ which explains the large spin polarisation. For negative bias the spin-$\downarrow$ component becomes dominant. The quality factor, however, does not reach the same height as for spin-$\uparrow$ due to the different density of states for the two spin species in the ferromagnet. For large bias electronic transport proceeds mainly through the spin-symmetric background so that the quality factor of spin-filtering goes to zero. Therefore the capability for spin measurement is a combined effect of the asymmetric density of states in the ferromagnet and the splitting of the Kondo resonances by the exchange field $\delta_{split}$.\\
Depending on the voltage bias a specific spin direction may be tuned to contribute to charge transfer due to the splitting of the Kondo peak. Additionally, the Kondo peak defines an almost perfect interface as it alignes the interface spin with the bulk FM. This makes the F-QD-S setup an ideal spin filter. This is of special importance in the case of Cooper pair splitters\cite{PhysRevLett.104.026801,19829377,2009arXiv0910.5558W} where the final proof of entanglement heavily relies on an effective spin measurement.\cite{soller1,morten-2008-81,PhysRevLett.94.210601,PhysRevB.66.161320}\\
\section{Ferromagnet-Quantum Dot-Superconductor Device in the even state} \label{even}
\begin{figure}[ht]
\includegraphics[width=7cm]{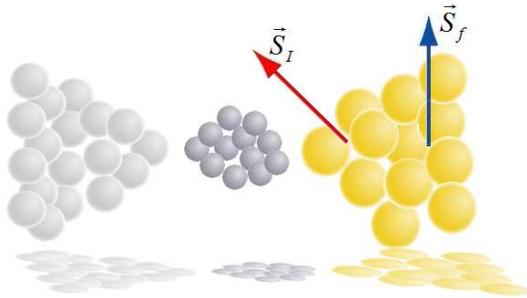}
\caption{The two magnetic moments of the bulk FM $\vec{S}_f$ and the interface $\vec{S}_I$ may be misaligned. This leads to spin-active scattering also in QD junctions.}
\label{fig6}
\end{figure}

Finally we want to investigate the even state of the QD. We did not need to incorporate spin-active scattering in the Kondo regime but with an even population of the QD the Kondo resonance dissapears. The absence of the collective state at the Fermi level of the FM allows for the possibility of interface effects. Indeed we find a pronounced mini-gap feature for an even charge state\cite{PhysRevLett.104.246804} which may be explained by a scenario based on spin-active scattering as in the case of a QPC considered in Section \ref{sfqpcsa} (see Fig. \ref{fig6}).
We consider an effective model for an interacting QD in an even charge state. Since the level spacing of the dot ($\Delta E \approx 0.4$meV) is significantly larger than the mini-gap energy we focus on a single orbital level in the discussion. The Hamiltonian for a simple resonant level coupled to a FM and a SC with spin-active scattering but still without the Coulomb interaction and the exchange field already has many parts given by
\begin{eqnarray}
H = H_f + H_s + H_d + H_{TRs} + H_{TRf} + H_{TR2}.
\end{eqnarray}
$H_f$ and $H_s$ again describe the FM and the SC, see Eqs. (\ref{hferro}) and (\ref{hsuper}). The QD, however, has to be taken into account explicitly now and is given by a resonant level at energy $\delta_\sigma$
\begin{eqnarray}      \label{16}
H_d = \sum_\sigma \delta_\sigma \, \tilde{d}_\sigma^+ \tilde{d}_\sigma \, ,
\end{eqnarray}
where $\tilde{d}_\sigma$ operator is the annihilator of the electron state on the dot and $\delta_\sigma = \delta + \sigma \Delta \epsilon/2$ is $\sigma$-dependent because of the exchange field $\Delta \epsilon$ induced by the bulk FM.
Eq.~(\ref{16}) is a good starting point since no collective state of the lead and the QD develops in the even charge state, which is characterised by $\delta = 0$. We first solve the resonant level case and then show how to implement the exchange field and Coulomb interaction in our effective model.

Because of the interface effects the magnetization axis on the ferromagnetic tunneling junction is rotated with respect to that in the bulk  as illustrated  in Fig.~\ref{fig6}. This is the essence of the spin-active scattering effect, see e.~g. [\onlinecite{PhysRevB.81.094508}].  In order to model that we include a spin-flip tunneling between the dot and the ferromagnet, thus the tunneling part of the Hamiltonian is given by
\begin{eqnarray}
H_{TR} &=& \sum_\sigma \tilde{\gamma}_s (\tilde{d}_\sigma^+ \Psi_{s\sigma} + h.c.) + \sum_\sigma \tilde{\gamma}_{f} (\tilde{d}_\sigma^+ \Psi_{f\sigma} + h.c.) \nonumber\\
&& + \sum_\sigma \tilde{\gamma}_{f2} (\tilde{d}_\sigma^+ \Psi_{f-\sigma} + h.c.). \label{htunnelr}
\end{eqnarray}
For computational reasons it is inconvenient to work with the spin-flip tunneling on the dot-FM interface. We choose to rotate the dot fields via
\begin{eqnarray}
d_\sigma = \frac{\gamma_{f} \tilde{d}_\sigma + \gamma_{f2} \tilde{d}_{-\sigma}}{\sqrt{\gamma_f^2 + \gamma_{f2}^2}}, \label{drot}
\end{eqnarray}
Rewriting Eq.~(\ref{htunnelr}) in the new basis we obtain
\begin{eqnarray*}
H_{TR\alpha} &=& \sum_{\sigma} \gamma_\alpha (d_\sigma^+ \Psi_{\alpha \sigma} + h.c.), \; \alpha = s,f\\
H_{TR2} &=& \sum_\sigma \gamma_{s2} (d_\sigma^+ \Psi_{s-\sigma} + h.c.), \; \mbox{where} \\
&& \gamma_{s} = \frac{\gamma_s \gamma_f}{\sqrt{\gamma_f^2 + \gamma_{f2}^2}}, \; \gamma_{s2} = \frac{\gamma_s \gamma_{f2}}{\sqrt{\gamma_f^2 + \gamma_{f2}^2}}. \label{htunnelr2} \, .
\end{eqnarray*}
So the spin-flip tunneling is effectively shifted to the dot-SC interface. Obviously, the above transformation does not change the dot Hamiltonian (\ref{16}). 

First, we consider the QD without the exchange field ($\delta_\sigma=0$) and onsite Coulomb interaction. In order to access the CGF, we need to introduce two counting fields for the separate leads. This means that now we have three contributions compared to Eq. (\ref{cgf2})
\begin{eqnarray}
\chi_{RSFa}(\lambda) = \langle T_{\cal C} \exp [-i \int_{\cal C} (T_R^{\lambda(t)} + T_{R2}^{\lambda(t)}+ T_{R3}^{\lambda(t)}) dt]\rangle
\end{eqnarray}
where $T_R^{\lambda(t)}$ and $T_{R2}^{\lambda(t)}$ represent $H_{TRs}$ and $H_{TR2}$ with the substitution $\Psi_{s\sigma}(x=0) \rightarrow \Psi_{s\sigma} (x=0) e^{-i \lambda_s(t)/2}$ and $T_{R3}^{\lambda(t)}$ can be obtained from $H_{TRf}$ with the substitution $\Psi_{f\sigma} (x=0) \rightarrow \Psi_{f\sigma} (x=0) e^{-i \lambda_f(t)/2}$. Using the Hamiltonian approach as before we arrive at the CGF given in Appendix \ref{appB}, Eq. (\ref{srcgf}).

The emerging formula is formally identical to the result for the SFQPC with spin-active scattering in Eq.~(\ref{spincgf}) but with energy-dependent transmission coefficients. Above the gap we observe single electron transmission and spin flip transmission while below the gap we obtain AR and SAR.
The spin-active scattering leads to triplet correlations in the ferromagnet. This kind of proximity phenomenon is mediated by the QD instead of a tunneling contact as in Section \ref{sfqpcsa}. The new feature in our setup is that the triplet correlations feel the exchange field of the bare ferromagnet which allows for a qualitatively new mini-gap feature.

\begin{figure}[ht]
\includegraphics[width=8cm]{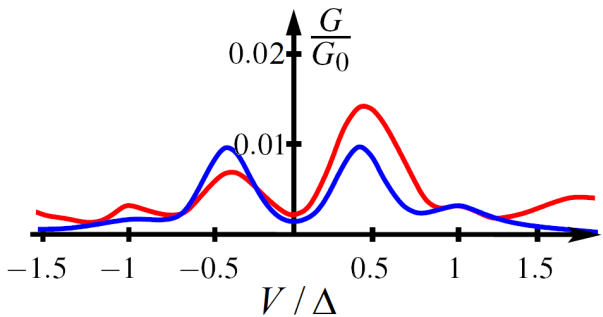}
	\caption{The conductance for a single channel contact with spin-active scattering is shown as a function of the voltage. The theoretical prediction by our model is the blue curve and the red curve refers to the experimental data taken from \cite{PhysRevLett.104.246804} at a background voltage $V_{BG} = 11.175 V$. The theoretical fit has been done using the parameters $\Gamma_f = 0.01\Delta, \; \Gamma_s = 0.005\Delta, \;\Gamma_{s2} = 0.015\Delta,\; P = 0.46, \; T=0.1\Delta, \; U= 2 \Delta$ and the gap $\Delta_0$ has been chosen such that the peaks are at the correct position $\Delta_0 = 0.9 \Delta$. Furthermore we can infer $\delta_\uparrow = 0.4 \Delta$.}
	\label{fig:fqs2a}
\end{figure}

Let us now turn to the situation of an interacting QD with a level splitting given by finite $\delta_\sigma$. The inclusion of the Coulomb interaction is done in the same way as in [\onlinecite{PhysRevB.68.035105}] for a Josephson junction: for the Andreev reflection transmission coefficients the second spin level has to account for the local exchange field $U$. This procedure is formally equivalent to a mean field solution including Coulomb interaction. The later analysis of the experimental data shows $\Delta \epsilon \lesssim \Delta_0$, $\Delta_0$ being much larger than the tunnel rates of the dot to the FM/SC lead. Using this assumption the result may be greatly simplfied: for energies above the gap spin-flip transmissions (and thus $H_{TR2}$) can be neglected since they involve both spin species. Below the gap AR involves both spin species and can therefore be neglected as well. The CGF for the resulting effective model for a QD in the even state is given by
\begin{widetext}
\begin{eqnarray}
\ln \chi_{es} (\lambda, \tau) &=& 2\tau \int \frac{d\omega}{2\pi} \left(\ln \{1+ \sum_\sigma T_{ese\sigma} [n_{f+} (1-n_s) (e^{i \lambda} -1) + n_s (1-n_{f+}) (e^{-i \lambda} -1)]\} \theta\left(\frac{|\omega| - \Delta_0}{\Delta_0}\right)\right. \nonumber\\
&& + \left. \frac{1}{2} \ln \{1+ \sum_\sigma T_{esAT\sigma} [n_{f+} (1-n_{f-}) (e^{2i \lambda} -1) + n_{f-} (1-n_{f+}) (e^{-2i \lambda} -1)]\} \theta \left(\frac{\Delta_0 - |\omega|}{\Delta_0}\right)\right), \label{escgf}
\end{eqnarray}
where the transmission coefficients are
\begin{eqnarray*}
T_{ese\sigma} &=& \frac{4 \Gamma_{f\sigma} \Gamma_{s11}}{(\Gamma_{f\sigma} + \Gamma_{s11})^2 + (\omega - \delta_\sigma)^2},\\
T_{esAT\sigma} &=& \frac{4\Gamma_{s23}^2 \Gamma_{f\sigma}}{(\omega - \delta_\sigma)^2(\omega - \delta_\sigma + U)^2 + (\Gamma_{s23}^2 + \Gamma_{f\sigma}^2)^2 + \Gamma_{s23}^2 (\omega - \delta_\sigma) (\omega - \delta_\sigma + U) + \Gamma_{f\sigma}^2(\omega - \delta_\sigma) (\omega - \delta_\sigma + U)},
\end{eqnarray*}
\end{widetext}
We have used the abbreviations $\Gamma_{f\sigma} = \Gamma_f (1 + \sigma P), \; \Gamma_{s11} = \Gamma_s |\omega| / \sqrt{|\omega^2 - \Delta_0^2|}$ and $\Gamma_{s23} = 2 \sqrt{\Gamma_s \Gamma_{s2}} \Delta_0 / \sqrt{|\Delta_0^2 - \omega^2|}$ that involve $\Gamma_f = \pi \rho_f \gamma_f^2 /2, \; \Gamma_s = \pi \rho_s \gamma_s^2/2$ and $\Gamma_{s2} = \pi \rho_s \gamma_{s2}^2/2$.
We compare the results of our model described above to the experimental data in Fig.~\ref{fig:fqs2a}.
We see that spin-active scattering in the presence of Coulomb interaction may lead to a significant mini-gap feature with a width of $\approx \Delta$ and conductance peaks even higher than the ones associated to the SC gap. The effective model correctly predicts the four-peak structure referring to the SC DOS and the exchange field as one observes in the experiment. It also explains the relation of the mini-gap feature to the ferromagnetic exchange field: SAR occurs via just a single spin level on the QD since it is associated to triplet correlation functions in the bare ferromagnet. In the presence of the ferromagnetic exchange field the two spin levels of the QD split. The exchange field therefore causes a splitting of the two SAR conductance peaks which is directly observable via the new mini-gap feature. This splitting is directly given by $\Delta \epsilon$.
We do not get the asymmetry inside the gap. This is due to the shortcomings of our model. An obvious improvement would be a more sophisticated (including correlation effects) calculation in the interaction.

Concerning the approximations we made to arrive at Eq.~(\ref{escgf}) we should note that indeed the exchange field $\Delta \epsilon \approx \Delta \gg \Gamma_{f,s,s2}$ so that Andreev reflection and single electron spin-flip transmission may be neglected.

In the experiment also the dependence of the subgap feature on an external magnetic field has been investigated. One observes that the subgap feature does almost not evolve in the magnetic field as long as the superconducting gap is not fully closed. If the gap closes, also the mini-gap feature gets strongly suppressed. We can analyse the evolution in magnetic field as well, using the model derived above. We just need to add the term $\sigma g \mu_B B/2$ to the positions of the split levels in Eq.~(\ref{escgf}), where $B$ refers to the external magnetic field. We use a typical value\cite{doi:10.1021/nl802418w} of $g=8$.\footnote{For the charge state analysed in Section \ref{kondo} we obtained $g=12$. The $g$-factors depend on the charge state considered and we assume it to be $g=8$ here.} The evolution of the superconducting gap is assumed to be given by
\begin{eqnarray*}
\Delta_0(B) = \Delta_0 \sqrt{1- \left(\frac{B}{B_c}\right)^2},
\end{eqnarray*}
and $\Delta_0$ is substituted in Eq. (\ref{escgf}) with $\Delta_0(B)$. For $\Delta_0(B) = 0$ the effective model given by Eq. (\ref{escgf}) should not be applicable anymore but since we are only interested in the evolution below the gap we assume the conductance for $\Delta_0(B) = 0$ to be constant for the voltage range considered here.
\begin{figure}[ht]
\includegraphics[width=9cm]{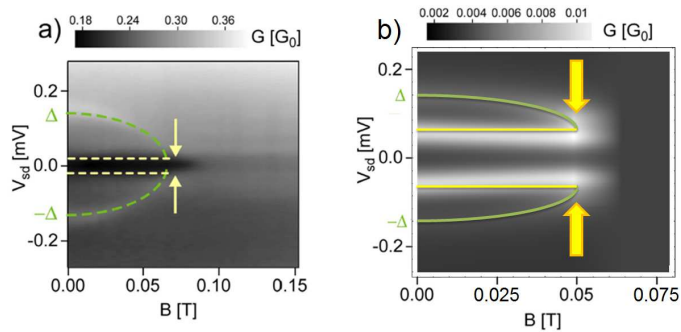}
	\caption{The evolution of the subgap feature in magnetic field is shown: a) experimental data for a typical sample different from the one analysed in Fig. \ref{fig:fqs2a}, (different charge state). One observes a subgap feature at the energy scale of the exchange field indicated in yellow. The feature is suppressed above the critical field of the SC (indicated by yellow arrows). b) shows the conductance as a function of magnetic field given by Eq. (\ref{escgf}). We assume $g=8$ and for vanishing superconducting gap we assume the conductance to be constant for the voltage range considered here. The critical magnetic field is taken to be $B_c = 64$mT and $\Delta_0 = 0.14$meV.}
	\label{fig:bfield}
\end{figure}
We find that our model correctly predicts the qualitative behavior of the mini-gap feature. The gap closes whereas the mini-gap stays in place as long as the gap is not vanishing. This is related to the very large exchange field observed in the experiment. $\delta_\uparrow = 0.4 \Delta$ corresponds to a critical magnetic field of the subgap feature of $B_{c,\mbox{subgap}} = 0.19$T, which is much larger than the critical magnetic field of the SC. Therefore we conclude that our model delivers a qualitatively correct description of the underlying physics. According to our explanation the splitting of the SAR peaks due to the exchange field gives direct evidence for the triplet correlations due to spin-active scattering since normal (spin-symmetric) AR conductance peaks cannot split up in an applied magnetic field. In this way the spin-active scattering can be identified in simple transport experiments in a way similar to the explicit investigation of Andreev bound states.\cite{2010arXiv1012.3867H}\\

\section{Conclusion} \label{conclusion}
Concluding, we have calculated the FCS for SFQPCs with and without spin-active scattering at the interface. We have demonstrated the necessity to take it into account for a consistent explanation of the current-voltage characteristics. Using these results we derived an effective description of a F-QD-S contact in the Kondo regime. In this case the Kondo effect imposes a strong asymmetry between the coupling of the SC and the FM to the QD. Spin-active scattering at the interface is strongly suppressed making the device an ideal tool for spin measurements in Cooper pair splitters. Spin-active scattering may be 'switched on' in an even charge state of the QD. There, as in the case of SFQPCs, it induces triplet correlations that lead to an observable mini-gap feature. Furthermore, our model allows to reproduce and interpret the evolution of the mini-gap feature in an external magnetic field.\\
HS would like to thank S. Maier, K. F. Albrecht and D. Breyel for many interesting discussions. The financial support was provided by the DFG under grant No. KO--2235/3, by the Kompetenznetz "Funktionelle Nanostrukturen III" of the Baden-W\"{u}rttemberg Stiftung (Germany), the EU FP7 project SE$^2$ND, EU ERC CooPairEnt 258789, OTKA CNK80991 and TAMOP 4.2.1./B-09/1/KMR-2010-0002. S. C. acknowledges support by the Bolyai Janos Scholarship.
\appendix
\begin{widetext}
\section{Expression for the CGF for the SFQPC with spin-active scattering} \label{appA}
The CGF for a SC-FM-quantum point contact is quite complicated in full detail and shall not be reported here. To give a clear physical understanding of the processes involved we use an approximation for the non-interacting self-energy due to the superconducting lead that is also used e.g. in [\onlinecite{Raimondi19991141}]: we treat the non-interacting self-energy of the SC to be real and purely off-diagonal for energies below the superconducting gap and to be diagonal for energies above $\Delta_0$. The one due to the normal-lead is always diagonal. This approximation is valid for $V \ll \Delta_0$ as well as $V \gg \Delta_0$ so that still all relevant charge transfer processes are included. For the plots and the comparison to experimental data we use, of course, the full model, which gives a slightly different behavior around $V=\Delta_0$. Using this approximation of the tunneling self-energy we arrive at
\begin{eqnarray*}
\ln \chi_{SFa}(\lambda) &=& 2 \tau \int \frac{d\omega}{2\pi} [\ln (\{1+ T_{e\sigma}[n_{f+}(1-n_s) (e^{i \lambda} -1) + n_s(1-n_{f+}) (e^{-i \lambda} -1)]\} \nonumber\\
&& \{1+ T_{e-\sigma}[n_{f+}(1-n_s) (e^{i \lambda} -1) + n_s(1-n_{f+}) (e^{-i \lambda} -1)]\} \nonumber\\
&& - T_d[n_{f+}(1-n_s) (e^{i \lambda} -1) + n_s(1-n_{f+}) (e^{-i \lambda} -1)]^2 \nonumber\\
&& - T_s[n_{f+}(1-n_s) (e^{i \lambda}-1) + n_s(1-n_{f+}) (e^{-i \lambda} -1)])\theta[(|\omega| - \Delta_0)/\Delta_0]\nonumber
\end{eqnarray*}
\begin{eqnarray}
&& + 1/2 \ln (\{1+ T_A[n_{f+}(1-n_{f-}) (e^{2i \lambda} -1) + n_{f-}(1- n_{f+}) (e^{-2i \lambda} -1)]\}^2 \nonumber\\
&& - T_{A2} [n_{f+}(1-n_{f-}) (e^{2i \lambda} -1) + n_{f-}(1- n_{f+}) (e^{-2i \lambda} -1)] \nonumber\\
&& + T_{AT} [n_{f+}(1-n_{f-})(e^{2i \lambda} -1) + n_{f-}(1-n_{f+}) (e^{-2i \lambda} -1)])\theta[(\Delta_0 - |\omega|)/\Delta_0]] \label{spincgf}.
\end{eqnarray}
We define the transmission coefficients to be
\begin{eqnarray*}
T_{e\sigma} &=& \frac{4(\beta_{11\sigma} + \beta_{12\sigma})}{(1+ \beta_{11\uparrow} + \beta_{12\uparrow})(1+ \beta_{11\downarrow} + \beta_{12\downarrow}) - \beta_{13\uparrow} \beta_{13\downarrow}}, \; T_d = \frac{16 \beta_{13\uparrow} \beta_{13\downarrow}}{[(1+ \beta_{11\uparrow} + \beta_{12\uparrow})(1+ \beta_{11\downarrow} + \beta_{12\downarrow}) - \beta_{13\uparrow} \beta_{13\downarrow}]^2},\\
T_s &=& \frac{4 [(\beta_{11\uparrow} - \beta_{11\downarrow} + \beta_{12\uparrow} - \beta_{12\downarrow})^2 + \beta_{13\uparrow} \beta_{13\downarrow}]}{[(1+ \beta_{11\uparrow} + \beta_{12\uparrow})(1+ \beta_{11\downarrow} + \beta_{12\downarrow}) - \beta_{13\uparrow} \beta_{13\downarrow}]^2}, \; T_A = \frac{4[(\beta_{21\downarrow} + \beta_{22\downarrow})(\beta_{21\uparrow} + \beta_{22\uparrow}) - \beta_{23\uparrow} \beta_{23\downarrow}]}{W},\\
T_{A2} &=& \frac{4(\beta_{23\uparrow} + \beta_{23\downarrow})^2(\beta_{21\uparrow} + \beta_{22\uparrow})(\beta_{21\downarrow} + \beta_{22\downarrow})}{W^2}, \; T_{AT} = \frac{4(\beta_{23\uparrow} + \beta_{23\downarrow})^2[1 + \beta_{23\uparrow} \beta_{23\downarrow}]^2}{W^2}, \\
&& \mbox{where}\\
W &=& 1 + (\beta_{21\uparrow} + \beta_{22\uparrow})(\beta_{21\downarrow} + \beta_{22\downarrow}) [2 + (\beta_{21\uparrow} + \beta_{22\uparrow})(\beta_{21\downarrow} + \beta_{22\downarrow})] \\
&& - 2 (\beta_{21\uparrow} + \beta_{22\uparrow})(\beta_{21\downarrow} + \beta_{22\downarrow})\beta_{23\uparrow} \beta_{23\downarrow} + (1+ \beta_{23\uparrow})^2 (1 + \beta_{23\downarrow})^2.
\end{eqnarray*}
In these definitions we used the abbreviations
\begin{eqnarray*}
\beta_{11\sigma} &=& \frac{\beta_{n} (1+ \sigma P)|\omega|}{\sqrt{\omega^2 - \Delta_0^2}}, \; \beta_{12\sigma} = \frac{\beta_{f} (1+ \sigma P)|\omega|}{\sqrt{\omega^2 - \Delta_0^2}}, \;\beta_{13\sigma} = \frac{2 (1+ \sigma P) (\beta_{n}  \beta_{f})^{1/2} |\omega|}{\sqrt{\omega^2 - \Delta_0^2}}, \\
\beta_{21\sigma} &=& \frac{\beta_{n} (1+ \sigma P) \Delta_0}{\sqrt{\Delta_0^2 - \omega^2}}, \; \beta_{22\sigma} = \frac{\beta_{f} (1+ \sigma P) \Delta_0}{\sqrt{\Delta_0^2 - \omega^2}}, \; \beta_{23\sigma} = \frac{2 (1+ \sigma P) (\beta_{n}  \beta_{f})^{1/2} \Delta_0}{\sqrt{\Delta_0^2 - \omega^2}},
\end{eqnarray*}
with $\beta_{n} = \rho_{f} \rho_{0s} \gamma^2 \pi^2 /2$ and $\beta_{f} = \rho_f \rho_{0s} \gamma_2^2 \pi^2 /2$.
We observe a more complicated structure of the CGF compared to Eq. (\ref{sfcgf}). Above the gap we observe single-electron transmission for the different spins described by $T_{e\sigma}$. Additionally spin-flip transmission of single electrons must contribute giving rise to the transmission coefficients $T_d$ and $T_s$. In the numerator of $T_s$ there are two contributions since one additionally has to keep track of the asymmetric DOS for the different spins. Below the gap we find two types of Andreev reflection: $T_A$ and $T_{A2}$ describe the normal, spin-symmetric Andreev reflection (AR) and $T_{AT}$ describes spin-flip Andreev reflection (SAR).
\section{Expression for the CGF for the F-QD-S junction with spin-active scattering} \label{appB}
We use the same approximation of the self-energy as in Appendix \ref{appA} to simplify the expression. In this approximation the CGF for a F-QD-S junction may be expressed as
\begin{eqnarray}
\ln \chi_{RSFa}(\lambda) &=& 2 \tau \int \frac{d\omega}{2\pi} \big[\big(\ln \{1+ \left[\sum_\sigma T_{Re\sigma}\right] [n_{f+} (1-n_s) (e^{i \lambda} -1) + n_s (1- n_{f+}) (e^{i \lambda}-1)] \nonumber\\
&& + T_{Rd} [n_{f+} (1-n_s) (e^{i \lambda} -1) + n_s (1- n_{f+}) (e^{i \lambda}-1)]^2\nonumber\\
&& - T_{Rs} [n_{f+} (1-n_s) (e^{i \lambda} -1) + n_s (1- n_{f+}) (e^{i \lambda}-1)]\}\big) \theta\left(\frac{|\omega| - \Delta_0}{\Delta_0}\right)\nonumber\\
&& + \frac{1}{2} \left(\ln \{1+ 2 T_{RA}[(e^{2i \lambda}-1) n_{f+} (1-n_{f-}) + (e^{2i \lambda}-1) n_{f-} (1-n_{f+})]\right. \nonumber\\
&& + T_{RAd} [(e^{2i \lambda}-1) n_{f+} (1-n_{f-}) + (e^{2i \lambda}-1) n_{f-} (1-n_{f+})]^2\nonumber\\
&& \left. + (T_{RAT} + T_{RA2}) [(e^{2i \lambda}-1) n_{f+} (1-n_{f-}) + (e^{2i \lambda}-1) n_{f-} (1-n_{f+})]\}\right)\big]\theta\left(\frac{\Delta_0 - |\omega|}{\Delta_0}\right), \label{srcgf}
\end{eqnarray}
where we set $\lambda_f - \lambda_s =: \lambda$ and we have the transmission coefficients
\begin{eqnarray*}
T_{Re\sigma} &=& \frac{4 \Gamma_{f\sigma} (\Gamma_{s11} + \Gamma_{s12}) [(\Gamma_{f-\sigma} + \Gamma_{s11} + \Gamma_{s12})^2 + (\omega - \delta)^2 - \Gamma_{s13}^2]}{\det A_{R10}}, \; T_{Rd} = \frac{16 \Gamma_{f\uparrow} \Gamma_{f\downarrow} [(\Gamma_{s11} + \Gamma_{s12})^2 - \Gamma_{s13}^2]}{\det A_{R10}},
\end{eqnarray*}
\begin{eqnarray*}
T_{Rs} &=& \frac{16 \Gamma_{f\uparrow} \Gamma_{f\downarrow}}{\det A_{R10}}, \; \det A_{R10} = [(\Gamma_{f\uparrow} + \Gamma_{s11} + \Gamma_{s12})^2 + \omega^2 - \Gamma_{s13}^2][(\Gamma_{f\downarrow} + \Gamma_{s11} + \Gamma_{s12})^2 + \omega^2 - \Gamma_{s13}^2] \nonumber\\
&& + \Gamma_{s13}^2 [(\Gamma_{f\uparrow} - \Gamma_{f\downarrow})^2 + 4\omega^2], \; T_{RA} = \frac{4 \Gamma_{f\uparrow} \Gamma_{f\downarrow} (\Gamma_{s21} + \Gamma_{s22})^2}{\det A_{R20}},\\
T_{RAd} &=& \frac{16 \Gamma_{f\uparrow}^2 \Gamma_{f\downarrow}^2 ((\Gamma_{s21} + \Gamma_{s22})^2 - \Gamma_{s23}^2)^2}{(\det A_{R20})^2}, \; T_{RAT} = \frac{4 \Gamma_{s23}^2[(\Gamma_{f\uparrow}^2 + \Gamma_{f\downarrow}^2)(\Gamma_{s23}^2 - (\Gamma_{s21} + \Gamma_{s22})^2 + \omega^2)^2]}{(\det A_{R20})^2}\\
T_{RA2} &=& \left\{4 \Gamma_{s23}^2\Gamma_{f\uparrow}^2 \Gamma_{f\downarrow}^2 (\Gamma_{f\uparrow}^2 + \Gamma_{f\downarrow}^2 + 4(\Gamma_{s23}^2 + \omega^2)) - 2 \Gamma_{f\uparrow} \Gamma_{f\downarrow} (\Gamma_{s21} + \Gamma_{s22})^2 (\Gamma_{f\uparrow}^2 + \Gamma_{f\downarrow}^2 \right. \nonumber\\
&& \left. + 4 (\Gamma_{s23}^2 + \omega^2))\right\} / \left\{(\det A_{R20})^2\right\}\\
\det A_{R20} &=& \omega^4+ [(\Gamma_{s21} + \Gamma_{s22})^2 + \Gamma_{f\uparrow} \Gamma_{f\downarrow} - \Gamma_{s23}^2]^2 + 2 (\Gamma_{s21} + \Gamma_{s22})^2 \omega^2 \nonumber\\
&& + \Gamma_{f\uparrow}^2 \omega^2 + \Gamma_{f\downarrow} \omega^2 + \Gamma_{s23}^2 [(\Gamma_{f\uparrow} + \Gamma_{f\downarrow})^2 + 2\omega^2]\\
\end{eqnarray*}
Again we have used several abbreviations in these definitions
\begin{eqnarray*}
\Gamma_{f\sigma} &=& \Gamma_F (1+ \sigma P), \; \Gamma_{s11} = \frac{\Gamma_s |\omega|}{\sqrt{\omega^2- \Delta_0^2}},\; \Gamma_{s12} = \frac{\Gamma_{s2} |\omega|}{\sqrt{\omega^2 - \Delta_0^2}}, \; \Gamma_{s13} = \frac{2(\Gamma_s \Gamma_{s2})^{\frac{1}{2}} |\omega|}{\sqrt{\omega^2 - \Delta_0^2}}, \\
\Gamma_{s21} &=& \frac{\Gamma_s \Delta_0}{\sqrt{\Delta_0^2 - \omega^2}}, \; \Gamma_{s22} = \frac{\Gamma_{s2} \Delta_0}{\sqrt{\Delta_0^2 - \omega^2}}, \; \Gamma_{s23} = \frac{2(\Gamma_s \Gamma_{s2})^{\frac{1}{2}}\Delta_0}{\sqrt{\Delta_0^2 - \omega^2}}
\end{eqnarray*}
with $\Gamma_f = \pi \rho_f \gamma_f^2/2$, $\Gamma_s = \pi \rho_s \gamma_s^2/2$ and $\Gamma_{s2} = \pi \rho_s \gamma_{s2}^2 /2$.\\
The result may be interpreted as the result for the SF tunnel contact with spin-active scattering in Eq. (\ref{spincgf}). Above the gap $T_{Re\sigma}$ describes single electron transfer without spin flip and $T_{Rd}$ describes the consecutive transfer of two electrons with different spin (again without spin flip). $T_{Rs}$ refers to spin flip transmission of single electrons. Below the gap $T_{RA}$ and $T_{RA2}$ describe single Andreev reflection and $T_{RAd}$ describes two consecutive Andreev reflections intiated by electrons with opposite spin. Spin flip Andreev reflection is given by $T_{RAT}$.
\end{widetext}

\end{document}